\pdfoutput=1
\documentclass[preprint,showpacs,preprintnumbers,amsmath,amssymb,floatfix,endfloats*]{revtex4}

\usepackage{graphicx}
\usepackage{dcolumn}
\usepackage{bm}
\usepackage{amsfonts}

\DeclareMathAlphabet{\mathsfsl}{OT1}{cmr}{bx}{it}
\begin{document}
\title{Mechanical annealing of model glasses: Effects of strain amplitude and temperature}
\author{Nikolai V. Priezjev$^{1,2}$}
\affiliation{$^{1}$Department of Mechanical and Materials
Engineering, Wright State University, Dayton, OH 45435}
\affiliation{$^{2}$National Research University Higher School of
Economics, Moscow 101000, Russia}
\date{\today}
\begin{abstract}

Molecular dynamics simulations are performed to examine the dynamic
response of amorphous solids to oscillatory shear at finite
temperatures.  The data were collected from a poorly annealed binary
glass, which was deformed periodically in the elastic regime during
several hundred shear cycles. We found that the characteristic time
required to reach a steady state with a minimum potential energy is
longer at higher temperatures and larger strain amplitudes.  With
decreasing strain amplitude, the asymptotic value of the potential
energy increases but it remains lower than in quiescent samples. The
transient decay of the potential energy correlates well with a
gradual decrease in the volume occupied by atoms with large
nonaffine displacements. By contrast, the maximum amplitude of shear
stress oscillations is attained relatively quickly when a large part
of the system starts to deform reversibly.

\end{abstract}




\maketitle

\section{Introduction}

The mechanical properties of metallic glasses can be tailored by
designing and exploring various thermomechanical processing
routes~\cite{Egami13}.    It is generally accepted that in contrast
to crystalline materials where deformation under applied forces can
be understood via motion of topological line defects, the plastic
deformation in glassy systems involves highly localized
rearrangements of small groups of atoms~\cite{Argon79,Falk98}.
Atomistic simulation studies have shown that metallic glasses
typically fail by forming shear bands under either
monotonic~\cite{Falk06,Horbach16} or cyclic
loading~\cite{GaoNano15,Priezjev17,GaoSize17} conditions.  It was
recently demonstrated that during cyclic indentation loading even
within the nominal elastic regime, metallic glasses can be hardened
due to atomic-level structural changes that lead to stiffening in
regions of the otherwise preferred yielding
path~\cite{Schuh08,Schuh12,WangASS17}. However, the effects of
hardening and softening in metallic glasses during time-dependent
deformation protocol remain relatively unexplored.

\vskip 0.05in

In recent years, the dynamic response of disordered systems to
periodic shear strain deformation was extensively studied via
molecular dynamics
simulations~\cite{Priezjev13,Sastry13,Reichhardt13,Priezjev14,IdoNature15,Priezjev16,Priezjev16a,Kawasaki16,Sastry17,OHern17,Priezjev17,IdoAnt17,Zapperi17}
and experimental
measurements~\cite{Petekidis02,Fielding08,Chaikin08,Cipelletti14,Helgeson14,Manneville17}.
Notably, it was shown that during oscillatory athermal quasistatic
deformation below yield, amorphous solids evolve into periodic limit
cycles with reversible particle dynamics, while above the critical
strain amplitude, the system dynamics becomes chaotic with a
positive maximal Lyapunov exponent~\cite{Reichhardt13,IdoNature15}.
Interestingly, the characteristic time needed to reach a steady
state in periodically sheared particle suspensions diverges at the
critical strain amplitude~\cite{Chaikin08}. Furthermore, the
analysis of the avalanche statistics and percolation characteristics
in a model glass under periodic quasistatic loading suggests that
yielding is a first-order phase transition, and the asymptotic
potential energy per particle at strain amplitudes below the elastic
limit depends on the sample preparation
history~\cite{Sastry13,Sastry17}.    At finite temperatures, a
slowly annealed glass subjected to a small-amplitude periodic shear
remains in a state with a low potential energy; while near the
critical strain amplitude, after a number of shear cycles, the
formation of a shear band is detected, which is associated with a
sharp increase in the potential energy and with a distinct drop in
the amplitude of shear stress
oscillations~\cite{Priezjev16a,Priezjev17}. Nevertheless, the
effects of cooling rate, strain amplitude, period of oscillations
and temperature on the atomic structure and potential energy in
glassy systems remain not fully understood.

\vskip 0.05in

In this paper, the dynamic behavior of a model glass subjected to
slow oscillatory shear is investigated using molecular dynamics
simulations.  It will be shown that poorly annealed binary glasses,
periodically deformed below yield, attain states with a potential
energy that is lower (higher) than that in quiescent glasses
prepared with a fast (slow) cooling rate. Upon increasing strain
amplitude or temperature, the characteristic time to reach a steady
state increases from a few tens to hundreds of shear cycles.  It is
also found that particles with large nonaffine displacements
initially form large clusters whose sizes gradually decay with
increasing number of cycles until the potential energy levels off to
a constant value.

\vskip 0.05in

The rest of the paper is structured as follows. The description of
molecular dynamics simulations is provided in the next section. The
time series of the potential energy and shear stress at different
temperatures and strain amplitudes as well as the analysis of
nonaffine displacements are presented in Sec.\,\ref{sec:Results}. A
brief summary of the results is given in the last section.

\section{Molecular dynamics simulation model}
\label{sec:MD_Model}

The model system we study is a three-dimensional (80:20) binary
mixture originally developed by Kob and Andersen~\cite{KobAnd95} in
order to reproduce the properties of the metal alloy
$\text{Ni}_{80}\text{P}_{20}$~\cite{Weber85}.   In this model, the
interaction between two nearby atoms is described by the truncated
Lennard-Jones (LJ) potential:
\begin{equation}
V_{\alpha\beta}(r)=4\,\varepsilon_{\alpha\beta}\,\Big[\Big(\frac{\sigma_{\alpha\beta}}{r}\Big)^{12}\!-
\Big(\frac{\sigma_{\alpha\beta}}{r}\Big)^{6}\,\Big],
\label{Eq:LJ_KA}
\end{equation}
where the interaction parameters are set to $\varepsilon_{AA}=1.0$,
$\varepsilon_{AB}=1.5$, $\varepsilon_{BB}=0.5$, $\sigma_{AB}=0.8$,
$\sigma_{BB}=0.88$, and $m_{A}=m_{B}$~\cite{KobAnd95}. This
parametrization ensures that the system will not be crystallized at
low temperatures~\cite{KobAnd95}. For computational efficiency, the
cutoff radius is fixed
$r_{c,\,\alpha\beta}=2.5\,\sigma_{\alpha\beta}$.  In what follows,
the LJ units of length, mass, energy, and time are set to
$\sigma=\sigma_{AA}$, $m=m_{A}$, $\varepsilon=\varepsilon_{AA}$, and
$\tau=\sigma\sqrt{m/\varepsilon}$, respectively.   The Newton's
equations of motion were integrated numerically using the velocity
Verlet algorithm~\cite{Allen87,Lammps} with the time step $\triangle
t_{MD}=0.005\,\tau$.

\vskip 0.05in

\begin{figure}[t]
\includegraphics[width=10.cm,angle=0]{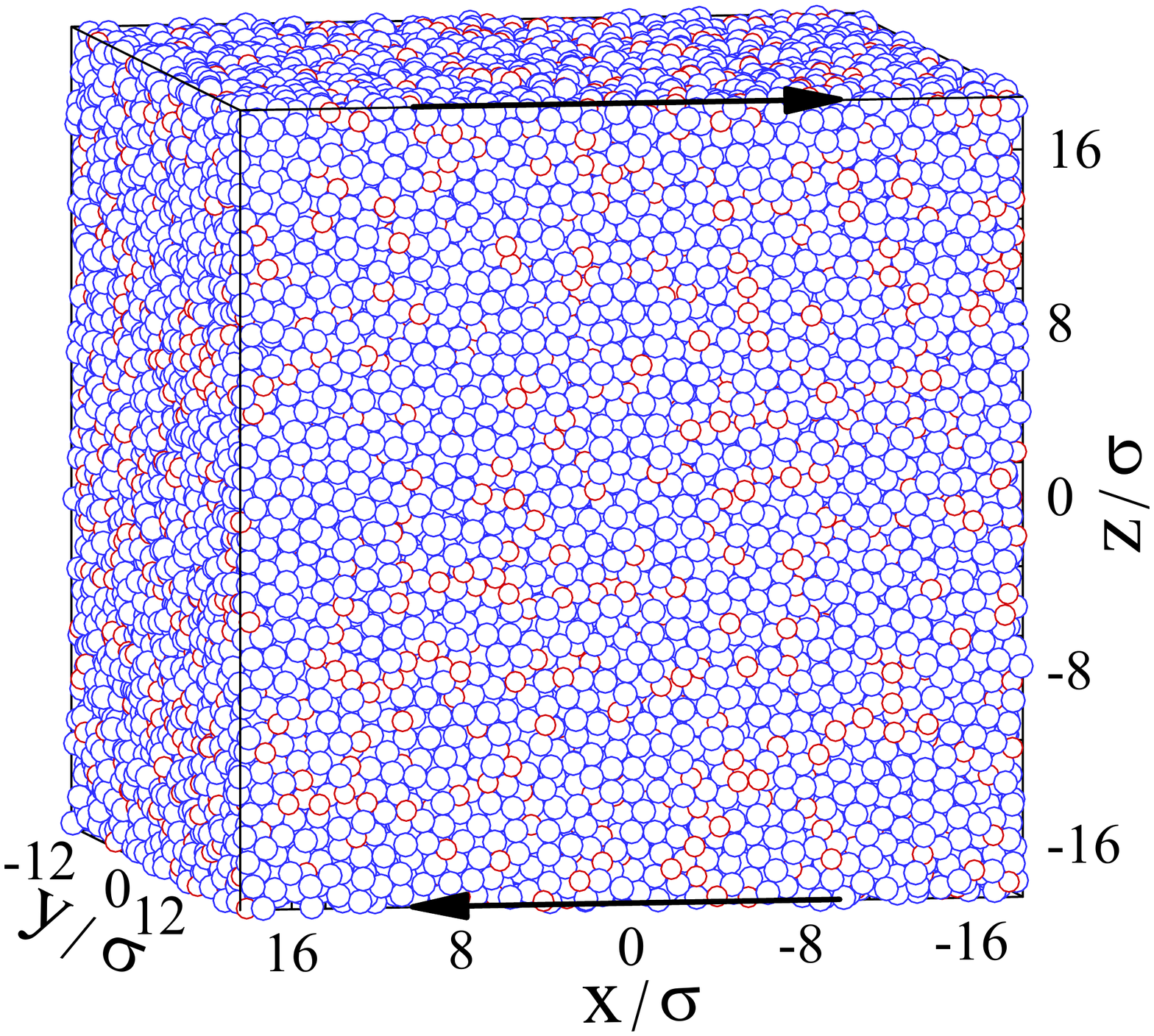}
\caption{(Color online) An instantaneous atom configuration of the
binary Lennard-Jones glass at the temperature
$T_{LJ}=10^{-2}\,\varepsilon/k_B$ after 600 shear cycles with the
strain amplitude $\gamma_0=0.06$ and oscillation period
$T=5000\,\tau$. The black arrows indicate the plane of shear. Two
types of atoms of different sizes are denoted by red and blue
circles. Atoms are not drawn to scale.}
\label{fig:snapshot_system}
\end{figure}

The system consists of $N=60\,000$ atoms that were placed in a
periodic box of linear size $L=36.84\,\sigma$ (see
Fig.\,\ref{fig:snapshot_system}). The MD simulations were performed
at a constant volume with the corresponding density
$\rho=\rho_{A}+\rho_{B}=1.2\,\sigma^{-3}$.  Initially, the system
was equilibrated in the absence of deformation at a high temperature
of $1.1\,\varepsilon/k_B$, which is above the critical temperature
$T_c\approx0.435\,\varepsilon/k_B$~\cite{KobAnd95}.  Throughout the
study, $k_B$ denotes the Boltzmann constant.   In our setup, the
temperature was regulated by the dissipative particle dynamics (DPD)
thermostat, which is known to stabilize the numerical integration of
the equations of motion and to avoid profile biasing in
nonequilibrium simulations~\cite{Soddemann03}.

\vskip 0.05in


In order to avoid the formation of a shear band at the initial stage
of the deformation protocol (described below), the system was
instantaneously quenched from a high temperature phase at
$T_{LJ}=1.1\,\varepsilon/k_B$ to a low temperature
$T_{LJ}=10^{-2}\,\varepsilon/k_B$ and kept undeformed at this
temperature during the time interval of $100\,\tau$.  Then, the
time-periodic shear strain deformation was applied along the $xz$
plane by using the Lees-Edwards periodic boundary conditions and the
SLLOD algorithm~\cite{Evans90} as follows:
\begin{equation}
\gamma(t)=\gamma_{0}\,\,\textrm{sin}(2\pi t / T),
\label{Eq:strain}
\end{equation}
where $\gamma_{0}$ is the strain amplitude and $T$ is the
oscillation period.    In this study, all simulations were performed
with the oscillation period $T=5000\,\tau$ and the oscillation
frequency $\omega=2\pi/T=1.26\times 10^{-3}\,\tau^{-1}$. During
production runs, atom positions were saved at the end of each cycle,
while the potential energy and shear stress were computed every
$5\,\tau$. Due to the relatively large system size and
implementation of the DPD thermostat, where stochastic forces have
to be generated for all pairs of nearby particles, the MD
simulations become computationally demanding.   For example, a
typical simulation of $40$ oscillation cycles using $32$ parallel
processors required about $110$ hours.   Thus, the data were
collected only in one sample at temperatures
$T_{LJ}\,k_B/\varepsilon=10^{-2}$, $10^{-3}$, $10^{-4}$, $10^{-5}$
and $10^{-6}$.

\section{Results}
\label{sec:Results}


It has long been realized that mechanical properties of glassy
materials depend crucially on the sample preparation
history~\cite{Schroers13}. In general, a slower thermal annealing
process allows for a thorough exploration of the potential energy
landscape, so that the system can find a deeper energy minimum, and,
as a result, upon external deformation, the dynamical yield stress
becomes larger~\cite{Procaccia13}. It was further shown that the
yield stress depends logarithmically on both the aging time and the
imposed shear rate~\cite{Varnik04}. In this study, we prepare poorly
annealed binary glass and periodically deform it below the yield
point at low temperatures. Under cyclic loading, the potential
energy landscape can be significantly distorted, which in turn might
promote plastic rearrangements, and upon reversal the system can be
relocated to deeper energy minima~\cite{Lacks04}. Although it is
intuitively expected that thermal fluctuations might facilitate
irreversible rearrangements of groups of atoms in a deformed glass;
however, the combined effect of temperature and strain amplitude on
the annealing process remains difficult to predict.

\vskip 0.05in

\begin{figure}[t]
\includegraphics[width=12.cm,angle=0]{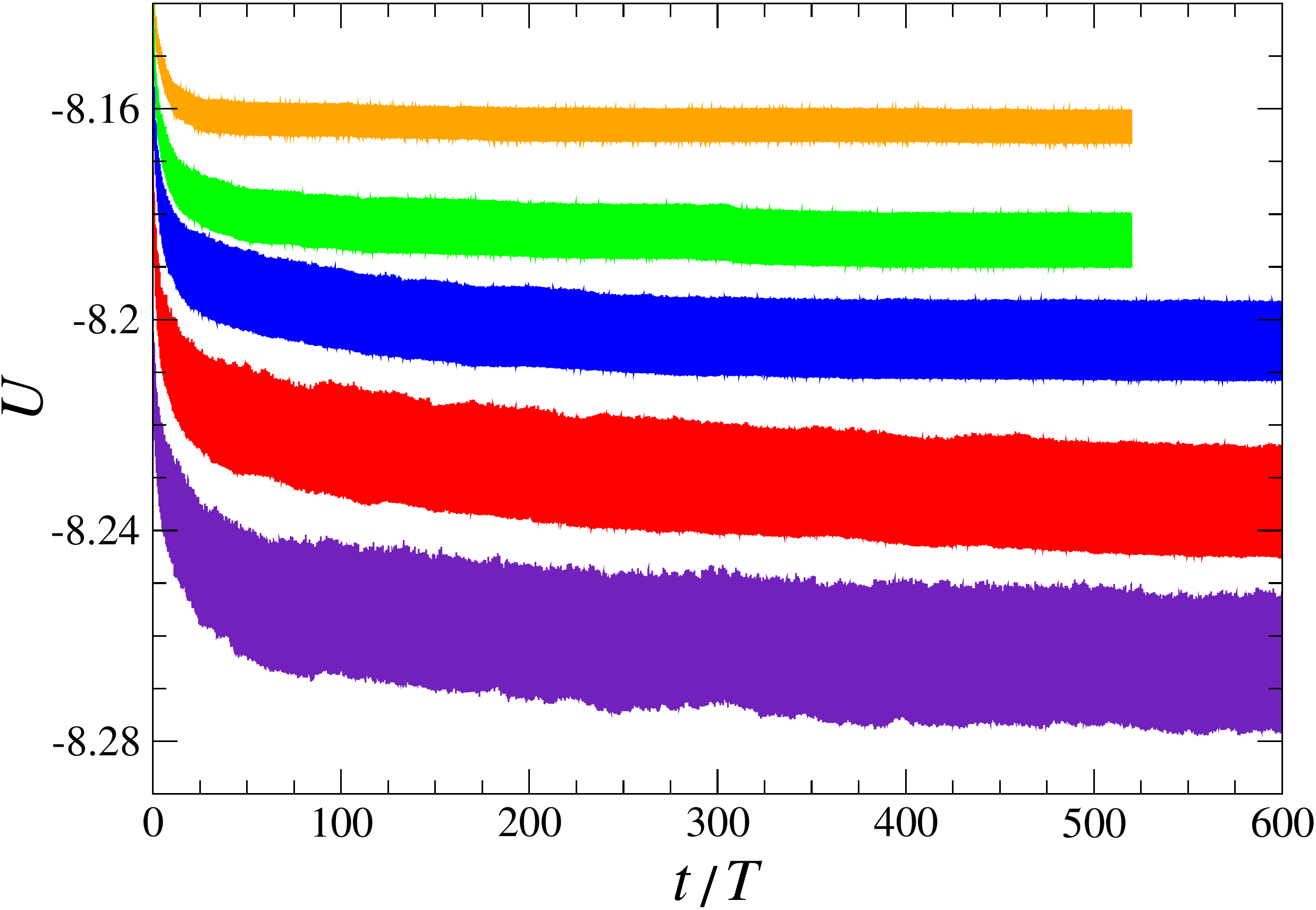}
\caption{(Color online) The variation of the potential energy per
particle $U$ (in units of $\varepsilon$) at
$T_{LJ}=10^{-2}\,\varepsilon/k_B$ during 600 oscillation cycles for
the strain amplitudes $\gamma_{0} = 0.03$, $0.04$, $0.05$, $0.06$,
and $0.07$ (from top to bottom).  For clarity, the data were
displaced vertically by $0.07\,\varepsilon$ for $\gamma_{0} = 0.03$,
by $0.06\,\varepsilon$ for $\gamma_{0} = 0.04$, by
$0.05\,\varepsilon$ for $\gamma_{0} = 0.05$, and by
$0.03\,\varepsilon$ for $\gamma_{0} = 0.06$. The oscillation period
is $T=5000\,\tau$. }
\label{fig:poten_time_T01}
\end{figure}

The time series of the potential energy per particle are presented
in Fig.\,\ref{fig:poten_time_T01} during 600 shear cycles for the
selected strain amplitudes and temperature
$T_{LJ}=10^{-2}\,\varepsilon/k_B$. It can be observed that at each
strain amplitude the minimum of the potential energy gradually
decreases with increasing number of cycles and then levels off to
constant values for $\gamma_{0} \leqslant 0.05$. Moreover, the time
interval needed to reach steady state becomes larger with increasing
strain amplitude. However, at finite temperatures it is difficult to
determine the exact number of shear cycles in the transient regime.
For example, one can notice in Fig.\,\ref{fig:poten_time_T01} that
at the strain amplitude $\gamma_{0} = 0.04$, the potential energy
nearly saturates to a constant value after about 200 cycles, which
is followed by a small drop in the potential energy at about 300
cycles.   It is also evident from Fig.\,\ref{fig:poten_time_T01}
that at large strain amplitudes, $\gamma_{0} = 0.06$ and $0.07$, the
transient regime exceeds 600 cycles.

\vskip 0.05in

\begin{figure}[t]
\includegraphics[width=12.cm,angle=0]{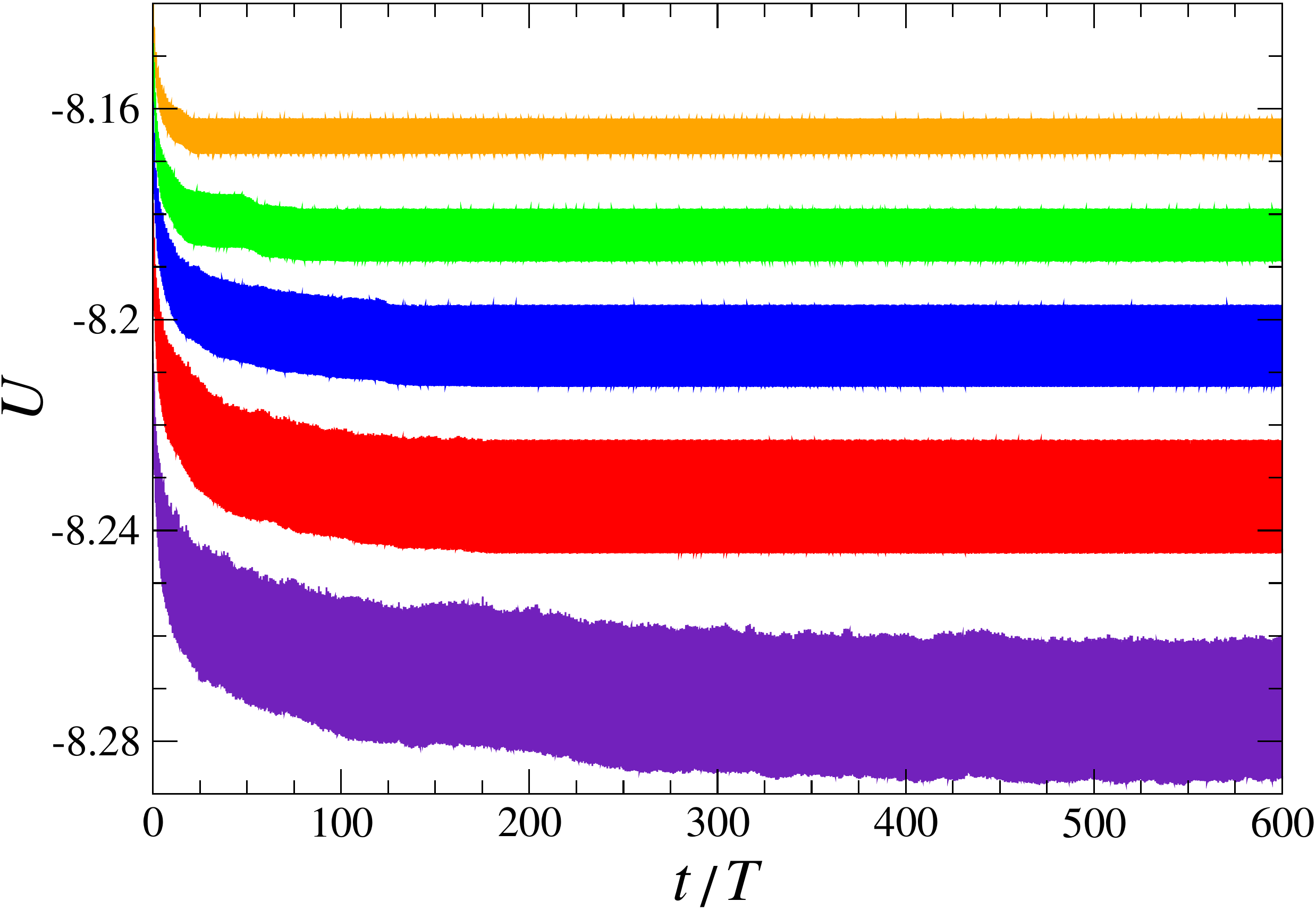}
\caption{(Color online) The time dependence of the potential energy
per particle $U$ (in units of $\varepsilon$) at the temperature
$T_{LJ}=10^{-5}\,\varepsilon/k_B$ during 600 oscillation cycles with
the period $T=5000\,\tau$ for $\gamma_{0} = 0.03$, $0.04$, $0.05$,
$0.06$, and $0.07$ (from top to bottom).  As in
Fig.\,\ref{fig:poten_time_T01}, the data were shifted upward by
$0.07\,\varepsilon$ for $\gamma_{0} = 0.03$, by $0.06\,\varepsilon$
for $\gamma_{0} = 0.04$, by $0.05\,\varepsilon$ for $\gamma_{0} =
0.05$, and by $0.03\,\varepsilon$ for $\gamma_{0} = 0.06$.  }
\label{fig:poten_time_T00001}
\end{figure}

As illustrated in Fig.\,\ref{fig:poten_time_T00001}, similar trends
in the decay of the potential energy can be observed at the lower
temperature $T_{LJ}=10^{-5}\,\varepsilon/k_B$.  It can be seen that
at a given strain amplitude, the transient regime of oscillations is
shorter than in the case $T_{LJ}=10^{-2}\,\varepsilon/k_B$ reported
in Fig.\,\ref{fig:poten_time_T01}.   Based on the results for
$T_{LJ} \leqslant 10^{-2}\,\varepsilon/k_B$, we thus conclude that
transition to the steady state regime of elastic deformation occurs
faster at lower temperatures.    We further comment that the minimum
value of the potential energy is lower for the strain amplitude
$\gamma_{0} = 0.07$ at $T_{LJ}=10^{-5}\,\varepsilon/k_B$ than for
$\gamma_{0} = 0.07$ at $T_{LJ}=10^{-2}\,\varepsilon/k_B$ (note that
the vertical scales are the same in Figs.\,\ref{fig:poten_time_T01}
and \ref{fig:poten_time_T00001}). Interestingly, the data shown for
$\gamma_{0} = 0.07$ in Fig.\,\ref{fig:poten_time_T00001} indicate
that the potential energy after about 400 cycles approaches a
constant value with superimposed fluctuations, which suggests the
presence of irreversible plastic events during periodic deformation.
By contrast, the minima of the potential energy at each cycle are
nearly the same in the steady regime at strain amplitudes
$\gamma_{0} \leqslant 0.06$, which is indicative of reversible
dynamics after each shear cycle (to be discussed below).    This
behavior is similar to that reported in athermal, quasistatic
oscillatory shear simulations of amorphous systems, which after some
time were found to settle into reversible limit
cycles~\cite{Reichhardt13,IdoNature15}.

\vskip 0.05in

\begin{figure}[t]
\includegraphics[width=12.cm,angle=0]{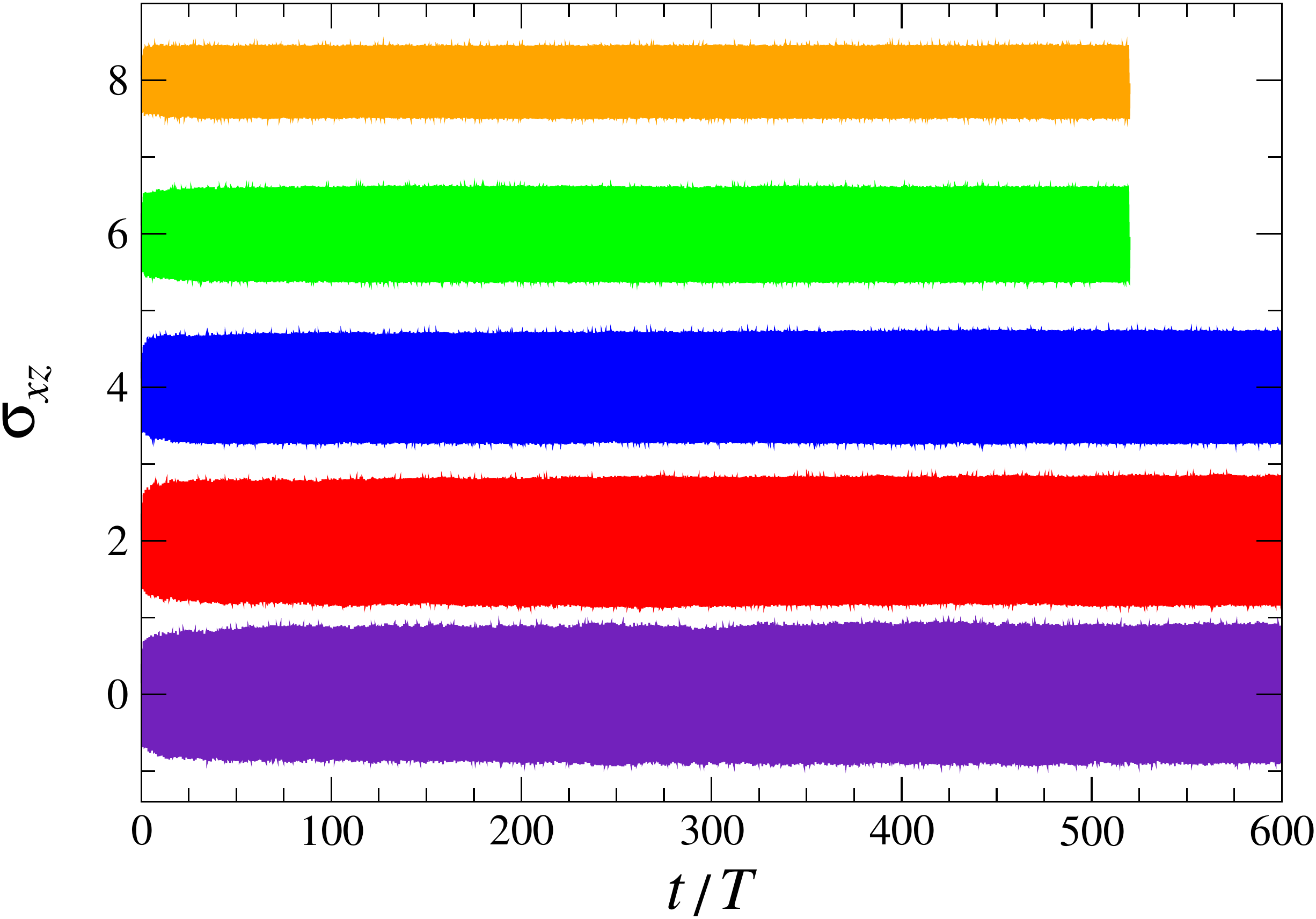}
\caption{(Color online) The shear stress $\sigma_{xz}$ (in units of
$\varepsilon\sigma^{-3}$) at the temperature
$T_{LJ}=10^{-2}\,\varepsilon/k_B$ for the strain amplitudes
$\gamma_{0} = 0.03$, $0.04$, $0.05$, $0.06$, and $0.07$ (from top to
bottom). For visualization, the data were displaced upward by
$8.0\,\varepsilon\sigma^{-3}$ for $\gamma_{0} = 0.03$, by
$6.0\,\varepsilon\sigma^{-3}$ for $\gamma_{0} = 0.04$, by
$4.0\,\varepsilon\sigma^{-3}$ for $\gamma_{0} = 0.05$, and by
$2.0\,\varepsilon\sigma^{-3}$ for $\gamma_{0} = 0.06$. The period of
oscillation is $T=5000\,\tau$. }
\label{fig:stress_strain_T01}
\end{figure}
\begin{figure}[t]
\includegraphics[width=12.cm,angle=0]{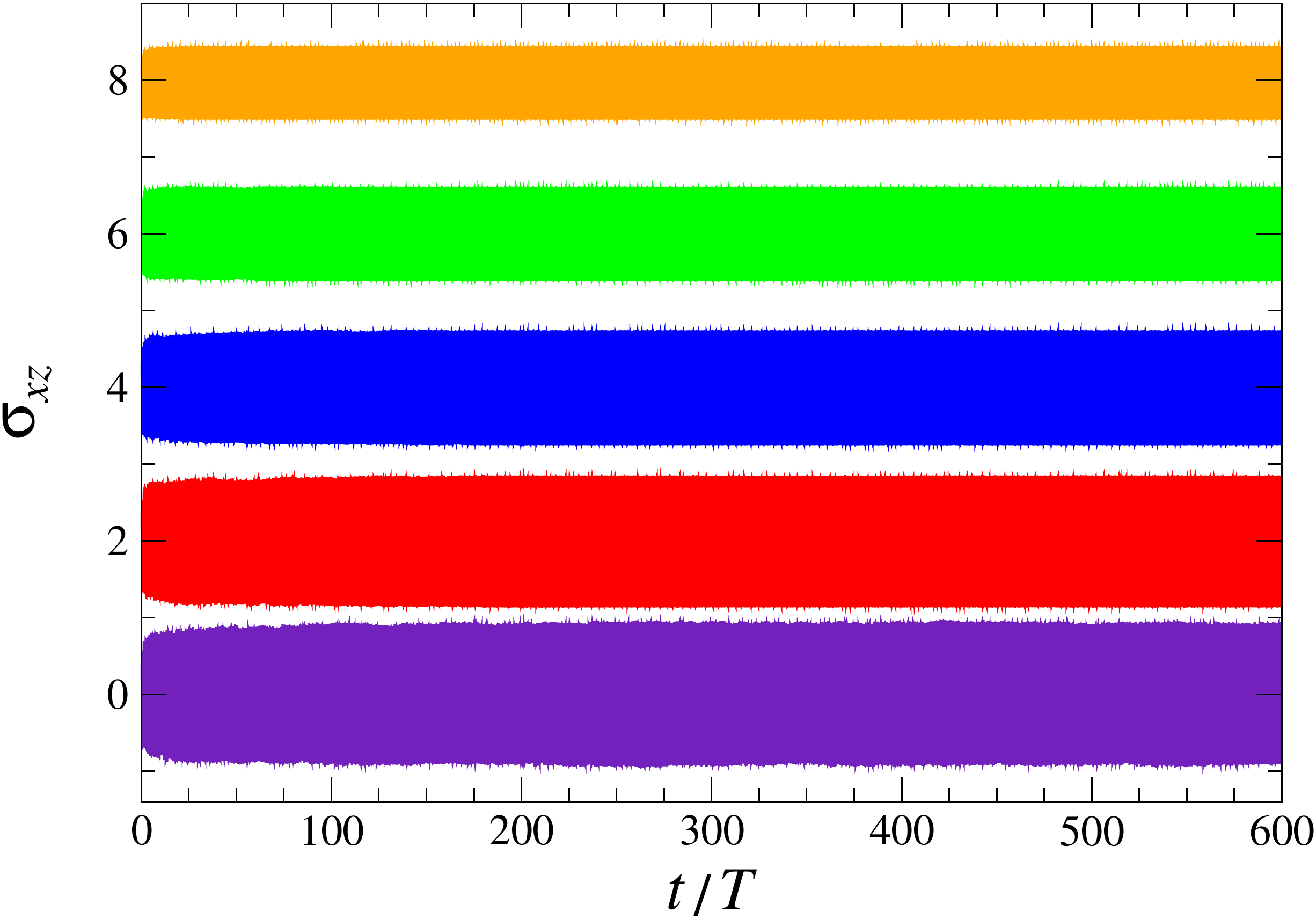}
\caption{(Color online) The dependence of shear stress $\sigma_{xz}$
(in units of $\varepsilon\sigma^{-3}$) at
$T_{LJ}=10^{-5}\,\varepsilon/k_B$ as a function of time for the
strain amplitudes $\gamma_{0} = 0.03$, $0.04$, $0.05$, $0.06$, and
$0.07$ (from top to bottom). Similar to
Fig.\,\ref{fig:stress_strain_T01}, the data were displaced by
$8.0\,\varepsilon\sigma^{-3}$ for $\gamma_{0} = 0.03$, by
$6.0\,\varepsilon\sigma^{-3}$ for $\gamma_{0} = 0.04$, by
$4.0\,\varepsilon\sigma^{-3}$ for $\gamma_{0} = 0.05$, and by
$2.0\,\varepsilon\sigma^{-3}$ for $\gamma_{0} = 0.06$. }
\label{fig:stress_strain_T00001}
\end{figure}

The variation of shear stress during 600 cycles at different strain
amplitudes are shown for $T_{LJ}=10^{-2}\,\varepsilon/k_B$ in
Fig.\,\ref{fig:stress_strain_T01} and for
$T_{LJ}=10^{-5}\,\varepsilon/k_B$ in
Fig.\,\ref{fig:stress_strain_T00001}.  It is clearly seen that the
amplitude of shear stress oscillations in the steady regime is
larger at higher strain amplitudes.   Somewhat surprisingly, we find
that the number of periods required to reach the maximum amplitude
of shear stress oscillations at a given $\gamma_0$ is significantly
smaller than the transient time for the variation of minima of the
potential energy reported in Figs.\,\ref{fig:poten_time_T01} and
\ref{fig:poten_time_T00001}. This difference is largest in the cases
$\gamma_{0} = 0.06$ and $0.07$ at $T_{LJ}=10^{-2}\,\varepsilon/k_B$,
where the amplitude of shear stress saturates after about 100
cycles, while the systems continue to explore states with
progressively lower potential energy minima during 600 cycles.

\vskip 0.05in

\begin{figure}[t]
\includegraphics[width=12.cm,angle=0]{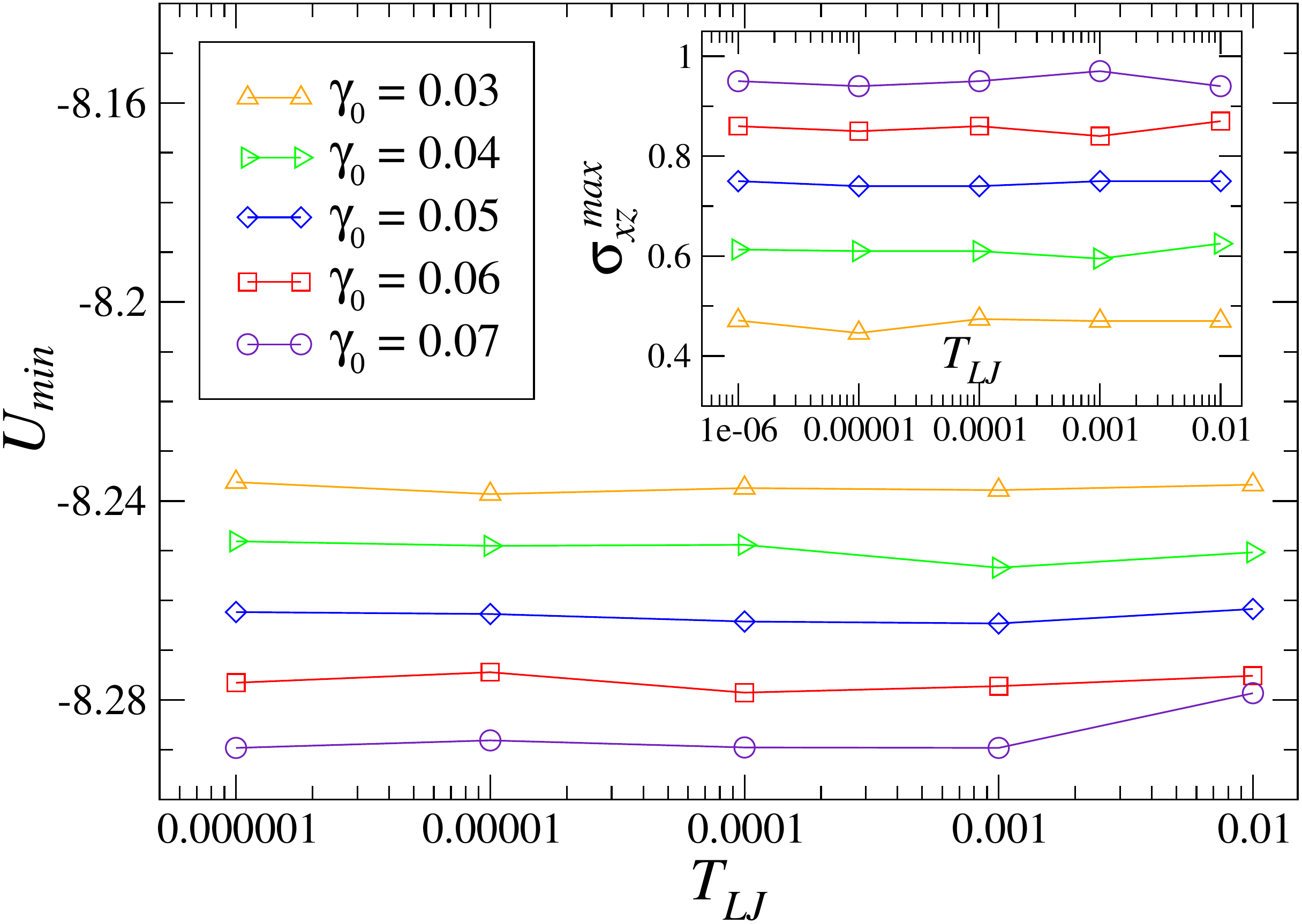}
\caption{(Color online) The minimum of the potential energy per
particle $U_{min}$ (in units of $\varepsilon$) as a function of
temperature for the indicated values of the strain amplitude
$\gamma_0$. The inset shows the maximum shear stress
$\sigma_{xz}^{max}$ (in units of $\varepsilon\sigma^{-3}$) for the
same strain amplitudes. }
\label{fig:min_poten_max_stress}
\end{figure}

The summary of the data for the maximum amplitude of shear stress
and minimum of the potential energy at different temperatures and
strain amplitudes are presented in
Fig.\,\ref{fig:min_poten_max_stress}. As is evident, both quantities
$U_{min}$ and $\sigma_{xz}^{max}$ are nearly independent of
temperature, except for the potential energy at the strain amplitude
$\gamma_{0} = 0.07$ and $T_{LJ}=10^{-2}\,\varepsilon/k_B$.  The
effect of cyclic loading on $U_{min}$ can be estimated by performing
an instantaneous quench from $T_{LJ}=1.1\,\varepsilon/k_B$ to a low
temperature in the range $10^{-5}\,\varepsilon/k_B \leqslant T_{LJ}
\leqslant 10^{-2}\,\varepsilon/k_B$ in the absence of shear. Such a
protocol yields a distinctly higher value of the potential energy,
$U\approx -8.21\,\varepsilon$, than those reported in
Fig.\,\ref{fig:min_poten_max_stress}.   On the other hand, it was
previously shown~\cite{Priezjev16a,Priezjev17} that a slow annealing
of an undeformed system across the glass transition with the rate of
$10^{-5}\,\varepsilon/k_B\tau$ to the target temperature
$T_{LJ}=10^{-2}\,\varepsilon/k_B$ results in the potential energy
$U\approx -8.31\,\varepsilon$, which is markedly lower than
$U_{min}$ for all $\gamma_{0}$ in
Fig.\,\ref{fig:min_poten_max_stress}. These results are consistent
with the conclusions obtained for the binary glass model using
athermal quasistatic oscillatory shear deformation in the elastic
range~\cite{Sastry17}.

\vskip 0.05in


Additional insight into the relaxation process under periodic
deformation can be gained through the analysis of the so-called
nonaffine displacements of atoms. To remind, the relative
displacements of atoms within a small group can be described by a
combination of a linear transformation and a deviation from a local
linear field~\cite{Falk98}.   Thus, the nonaffine measure can be
computed using the transformation matrix $\mathbf{J}_i$ that best
maps all vectors between the $i$-th atom and its neighbors during
the time interval $\Delta t$ according to:
\begin{equation}
D^2(t, \Delta t)=\frac{1}{N_i}\sum_{j=1}^{N_i}\Big\{
\mathbf{r}_{j}(t+\Delta t)-\mathbf{r}_{i}(t+\Delta t)-\mathbf{J}_i
\big[ \mathbf{r}_{j}(t) - \mathbf{r}_{i}(t)    \big] \Big\}^2,
\label{Eq:D2min}
\end{equation}
where the summation is performed over $N_i$ neighboring atoms within
the cutoff distance of $1.5\,\sigma$ from $\mathbf{r}_{i}(t)$.  Note
that, when normalized by the number of neighbors, the value $D^2
\approx 0.01\,\sigma^2$ is equal to the typical cage size, and,
therefore, nonaffine displacements with $D^2 \gtrsim 0.01\,\sigma^2$
correspond to cage breaking events, which in some cases can be
reversible~\cite{Priezjev16,Priezjev16a}. In particular, the
analysis of spatial configurations of atoms with large nonaffine
displacements allowed a clear identification of a system-spanning
shear band, which was spontaneously formed after a number of shear
cycles at the critical strain amplitude in a well-annealed binary
glass~\cite{Priezjev17}.

\vskip 0.05in

\begin{figure}[t]
\includegraphics[width=12.cm,angle=0]{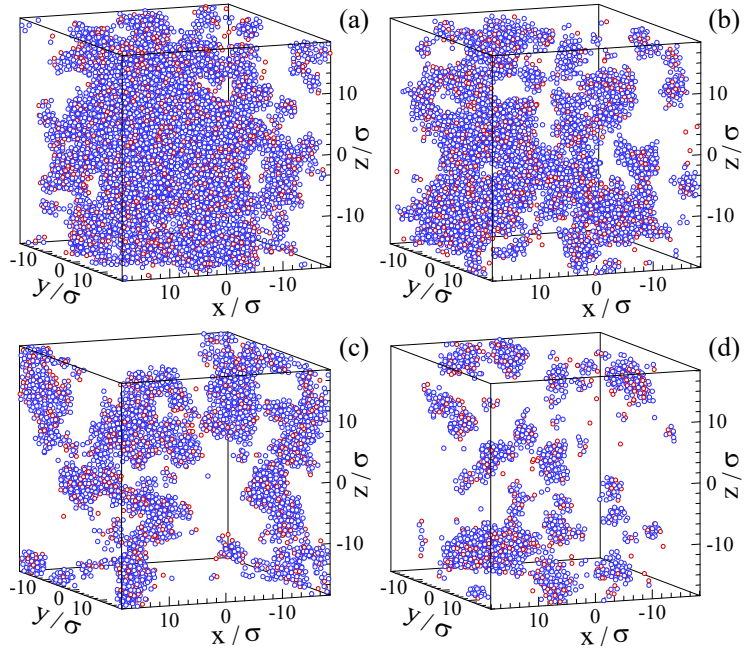}
\caption{(Color online) Snapshots of atom positions for the strain
amplitude $\gamma_{0}=0.06$, temperature
$T_{LJ}=10^{-2}\,\varepsilon/k_B$, and nonaffine measure (a)
$D^2(19T,T)>0.01\,\sigma^2$, (b) $D^2(79T,T)>0.01\,\sigma^2$, (c)
$D^2(199T,T)>0.01\,\sigma^2$, and (d) $D^2(599T,T)>0.01\,\sigma^2$.
}
\label{fig:snapshot_gam_T01_gam06}
\end{figure}
\begin{figure}[t]
\includegraphics[width=12.cm,angle=0]{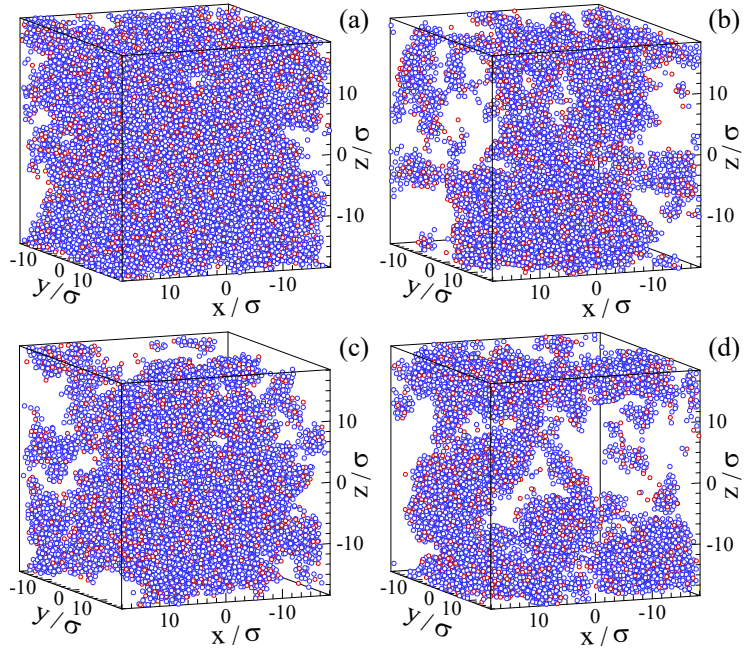}
\caption{(Color online) Spatial configurations of atoms for the
strain amplitude $\gamma_{0}=0.07$,
$T_{LJ}=10^{-2}\,\varepsilon/k_B$, and nonaffine measure (a)
$D^2(19T,T)>0.01\,\sigma^2$, (b) $D^2(79T,T)>0.01\,\sigma^2$, (c)
$D^2(199T,T)>0.01\,\sigma^2$, and (d) $D^2(599T,T)>0.01\,\sigma^2$.}
\label{fig:snapshot_gam_T01_gam07}
\end{figure}
\begin{figure}[t]
\includegraphics[width=12.cm,angle=0]{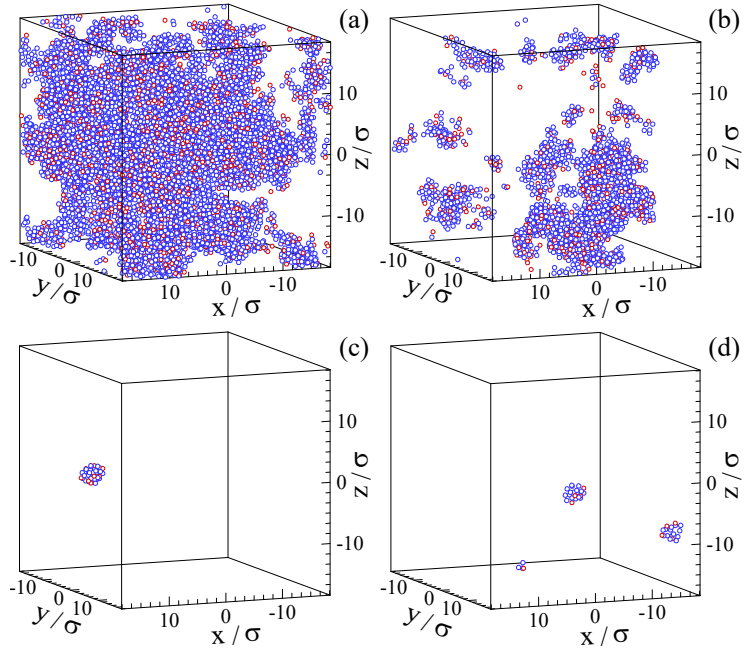}
\caption{(Color online) Atom positions for $\gamma_{0}=0.06$,
$T_{LJ}=10^{-5}\,\varepsilon/k_B$, and nonaffine measure (a)
$D^2(19T,T)>0.01\,\sigma^2$, (b) $D^2(79T,T)>0.01\,\sigma^2$, (c)
$D^2(199T,T)>0.01\,\sigma^2$, and (d) $D^2(599T,T)>0.01\,\sigma^2$.}
\label{fig:snapshot_gam_T00001_gam06}
\end{figure}
\begin{figure}[t]
\includegraphics[width=12.cm,angle=0]{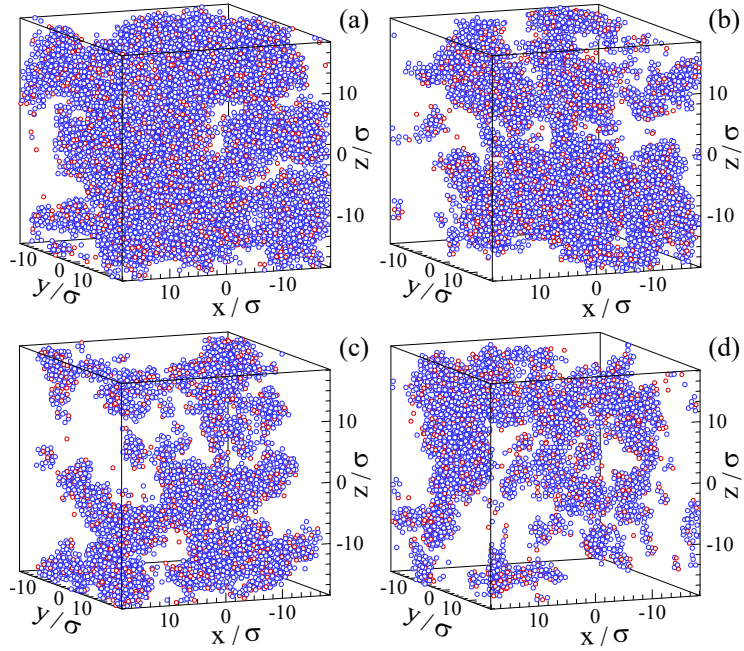}
\caption{(Color online) Instantaneous positions of atoms of types A
and B for the strain amplitude $\gamma_{0}=0.07$, temperature
$T_{LJ}=10^{-5}\,\varepsilon/k_B$, and nonaffine measure (a)
$D^2(19T,T)>0.01\,\sigma^2$, (b) $D^2(79T,T)>0.01\,\sigma^2$, (c)
$D^2(199T,T)>0.01\,\sigma^2$, and (d) $D^2(599T,T)>0.01\,\sigma^2$.}
\label{fig:snapshot_gam_T00001_gam07}
\end{figure}

In our analysis, the nonaffine measure was evaluated numerically
based on atom positions at the beginning and the end of each cycle
at zero global strain.  Typically, the quantity $D^2$ is broadly
distributed even at small strain amplitudes with the majority of
atoms undergoing reversible displacements with $D^2<0.01\,\sigma^2$
due to thermal
fluctuations~\cite{Priezjev16,Priezjev16a,Priezjev17}. Therefore, in
what follows, we only plot positions of atoms with
$D^2(nT,T)>0.01\,\sigma^2$, where $n$ is the cycle number. Four sets
of time sequences of spatial configurations of atoms with large
nonaffine displacements during a complete cycle are displayed in
Figs.\,\ref{fig:snapshot_gam_T01_gam06} and
\ref{fig:snapshot_gam_T01_gam07} for $\gamma_{0} = 0.06$ and $0.07$
at $T_{LJ}=10^{-2}\,\varepsilon/k_B$ and in
Figs.\,\ref{fig:snapshot_gam_T00001_gam06} and
\ref{fig:snapshot_gam_T00001_gam07} for $\gamma_{0} = 0.06$ and
$0.07$ at $T_{LJ}=10^{-5}\,\varepsilon/k_B$.    In each case, the
atoms with $D^2>0.01\,\sigma^2$ form clusters whose sizes gradually
decay with increasing number of cycles.  Notice that after $80$
cycles, the atoms with small nonaffine displacements (empty volume
in
Figs.\,\ref{fig:snapshot_gam_T01_gam06}-\ref{fig:snapshot_gam_T00001_gam07})
are organized into percolating regions that can deform elastically
and support relatively large shear stress.    This might explain why
the shear stress amplitude rather quickly approaches a constant
value, as shown in Figs.\,\ref{fig:stress_strain_T01} and
\ref{fig:stress_strain_T00001}.   At the same time, progressively
lower minima of the potential energy can still be attained via
small-scale plastic rearrangements leading to extended transients
shown in Figs.\,\ref{fig:poten_time_T01} and
\ref{fig:poten_time_T00001}.   We finally comment that at lower
strain amplitudes, $\gamma_{0} \leqslant 0.05$, the sequence of atom
configurations with large nonaffine displacements appears to be
qualitatively very similar to the one shown in
Fig.\,\ref{fig:snapshot_gam_T00001_gam06}, where large clusters
rapidly disappear during the first few tens of cycles followed by,
in some cases, a few isolated rearrangements in the steady regime of
periodic deformation.

\section{Conclusions}

In summary, large-scale molecular dynamics simulations were carried
out to investigate the influence of strain amplitude and temperature
on relaxation dynamics of a model glass. We considered a
three-dimensional binary mixture with highly nonadditive interaction
parameters that prevent crystallization at low temperatures and used
the dissipative particle dynamics thermostat that does not couple
particle dynamics to flow profile. The system was initially prepared
at a high temperature well above the glass transition, and then
following a fast quench to low temperature at constant volume, the
glass was subjected to periodic shear during hundreds of cycles.

\vskip 0.05in

It was found that at strain amplitudes below yield, the system
gradually evolves in a steady state with the potential energy that
is higher (lower) than that in a very slowly (quickly) annealed
glass in the absence of deformation. Moreover, with increasing
strain amplitude, the system is ultimately relocated to deeper
energy minima, and the characteristic time required to reach steady
state becomes longer at larger strain amplitudes and/or higher
temperatures.  Furthermore, the analysis of nonaffine displacements
indicates that after the first few tens of shear cycles, a large
part of the system starts to deform reversibly, which correlates
with the maximum amplitude of shear stress oscillations. On the
other hand, small-scale plastic rearrangements are associated with
extended transients before reaching steady state with a minimum
potential energy.

\section*{Acknowledgments}

Financial support from the National Science Foundation (CNS-1531923)
is gratefully acknowledged. The study has been in part funded by the
Russian Academic Excellence Project `5-100'. The molecular dynamics
simulations were performed using the efficient parallel code LAMMPS
developed at Sandia National Laboratories~\cite{Lammps}. This work
was supported in part by Michigan State University through
computational resources provided by the Institute for Cyber-Enabled
Research.


%

%

%

%

%


%

%

%

%

\bibliographystyle{prsty}

\end{document}